\documentclass[9pt,blue]{lapreprint}

\usepackage{lipsum}     
\usepackage[version=4]{mhchem} 
\usepackage{siunitx}    
\usepackage{pdflscape}  
\usepackage{rotating}   
\usepackage{textgreek}  
\usepackage{gensymb}    
\usepackage[misc]{ifsym} 
\usepackage{orcidlink}  
\usepackage{listings}   
\usepackage{colortbl}   
\usepackage{tabularx}   
\usepackage{longtable}  
\usepackage{subcaption}
\usepackage{multirow}
\usepackage{snotez}     
\usepackage{xcolor}
\DeclareSIUnit\Molar{M}

\title{Toxic comments reduce the activity of volunteer editors on Wikipedia}

\author[ \orcidlink{0000-0002-8347-6703} 1 \Letter]{Ivan Smirnov}
\author[2]{Camelia Oprea}
\author[1,3,4]{Markus Strohmaier}

\affil[1]{University of Mannheim}
\affil[2]{RWTH Aachen University}
\affil[3]{GESIS - Leibniz Institute for the Social Sciences}
\affil[4]{Complexity Science Hub Vienna}

\correspondence{smirnov@uni-mannheim.de}{IS}

\presentaddress[]{University of Mannheim, L15, 1–6, 68161 Mannheim, Germany}

\data[]{Data is available on \href{https://osf.io/2qyxj/}{OSF}.}

\compint[]{The authors declare no competing interests.}

\newcommand{\supplementarysection}{%
  \setcounter{figure}{0}
  \let\oldthefigure\thefigure
  \renewcommand{\thefigure}{S\oldthefigure}
  \setcounter{table}{0}
  \let\oldthetable\thetable
  \renewcommand{\thetable}{S\oldthetable}
}

\leadauthor{Smirnov}
\shorttitle{Impact of Toxicity on Wikipedia}

\begin{document}
\maketitle
\begin{abstract}
Wikipedia is one of the most successful collaborative projects in history. It is the largest encyclopedia ever created, with millions of users worldwide relying on it as the first source of information as well as for fact-checking and in-depth research. As Wikipedia relies solely on the efforts of its volunteer-editors, its success might be particularly affected by toxic speech. In this paper, we analyze all 57 million comments made on user talk pages of 8.5 million editors across the six most active language editions of Wikipedia to study the potential impact of toxicity on editors' behaviour. We find that toxic comments consistently reduce the activity of editors, leading to an estimated loss of 0.5--2 active days per user in the short term. This amounts to multiple human-years of lost productivity when considering the number of active contributors to Wikipedia. The effects of toxic comments are even greater in the long term, as they significantly increase the risk of editors leaving the project altogether. Using an agent-based model, we demonstrate that toxicity attacks on Wikipedia have the potential to impede the progress of the entire project. Our results underscore the importance of mitigating toxic speech on collaborative platforms such as Wikipedia to ensure their continued success.
\end{abstract}

\section{Introduction} \label{intro}
Wikipedia is arguably one of the most successful collaborative projects in history. It has become the largest and most-read reference work ever created, and it is currently the fifth most popular website on the Internet\footnote{\url{https://www.semrush.com/website/top/} [[Assessed on 23.02.2023]]}. Millions of users worldwide rely on Wikipedia as their first source of information when encountering a new topic, for fact-checking and in-depth research \cite{singer2017we}. Even if caution might be required when consulting less actively maintained pages \cite{bruckman2022should}, numerous studies have shown that Wikipedia is a reliable source of information in areas ranging from political science \cite{brown2011wikipedia} to pharmacology \cite{clauson2008scope} and its accuracy is comparable to traditional encyclopedias \cite{giles2005special} and textbooks \cite{kraenbring2014accuracy}.

One of the most remarkable aspects of Wikipedia's success is that its content is exclusively created and curated by volunteer-editors, known as Wikipedians. The English edition alone has 129,698 active editors\footnote{\url{https://en.wikipedia.org/wiki/List_of_Wikipedias} [[Assessed on 23.02.2023]]}. However, this volunteer-driven model also makes Wikipedia susceptible to the inherent challenges associated with maintaining such a large online community \cite{kraut2012building,keegan2017evolution}. For example, it has been previously observed that Wikipedia is not free of conflict, particularly in the form of so-called edit wars \cite{yasseri2012dynamics}, which impose significant costs on the project \cite{kittur2007he} and could negatively affect the quality of Wikipedia articles \cite{arazy2011information}.

In this paper, we focus on the impact of toxic comments directed towards editors on their activity. This aspect is less studied, but potentially not less important, as affected by toxic comments, Wikipedians might reduce their contributions or abandon the project altogether, threatening the success of the platform \cite{preece2001sociability}.

Toxicity has been extensively studied on popular social media websites such as Twitter \cite{chatzakou2017measuring,guberman2016quantifying}, Reddit \cite{xia2020exploring,almerekhi2020investigating}, and similar platforms \cite{wich2022introducing,silva2016analyzing}. However, much of these research focuses on automated toxicity detection and prevalence estimation rather than on evaluating its impact \cite{kiritchenko2021confronting}. As an online encyclopedia, Wikipedia is often perceived as immune to toxicity and has a strict “No personal attacks” policy \cite{no_personal}. Despite that, toxic speech and harassment have been previously observed on the platform \cite{arazy2013stay,harassment_survey,corple2016beyond,wulczyn2017ex,raish2019identifying}. The effects of such behaviors on editors’ contributions are, however, not well understood nor well studied. The largest study to date relies on a voluntary opt-in survey of the 3,845 Wikipedians conducted in 2015 \cite{harassment_survey}. It reports that 20\% of users witnessing harassment have stopped contributing for a while, 17\% considered not contributing anymore and 5\% stopped contributing at all.

In this paper, we analyzed all 57 million comments made on user talk pages of editors on the six most active language editions of Wikipedia (English, German, French, Spanish, Italian, Russian) to understand the impact of toxic speech on editors’ contributions (see \nameref{methods} for our definition of toxic comments). User talk pages are a place for editors to communicate with each other either on more personal topics or to extend their discussion from an article’s talk page. The majority of toxic comments are left on user talk pages \cite{qu2019wikidetox}. The comments we study were extracted from revision histories of talk pages and, thus, include even those toxic comments that were later archived or deleted by the page owner.

\begin{figure}
    \centering
    \includegraphics[width=0.8\hsize]{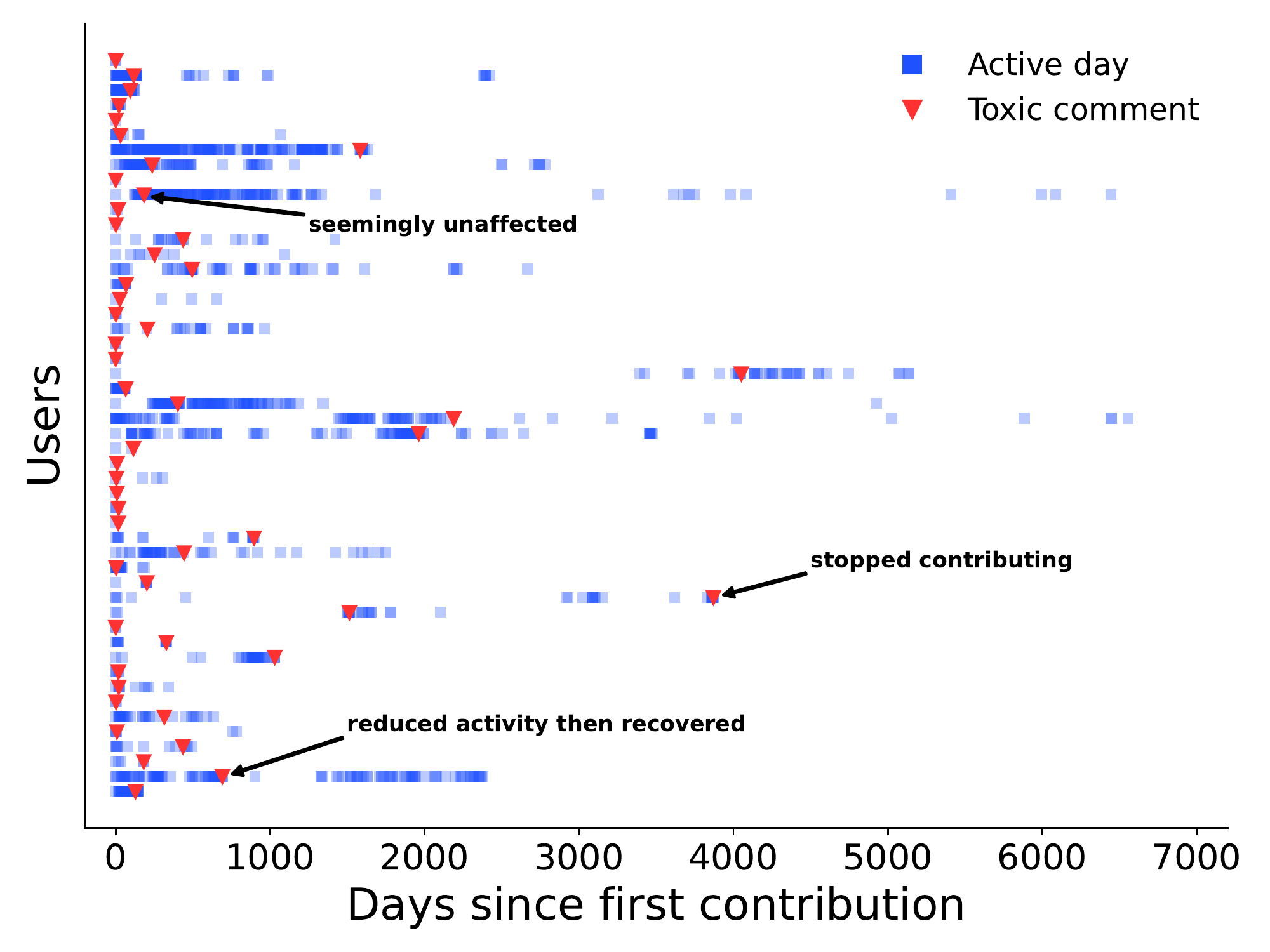}
    \caption{
        \textbf{After receiving a toxic comment many users temporarily reduce their activity or leave the project completely.
        }
        The figure shows the activity of 50 randomly selected users who received exactly one toxic comment. Blue squares indicate an active day, i.e. a day when at least one edit was done, starting from the first contribution of a given user. Red triangles correspond to toxic comments. Note that while some users are resilient and their activity is seemingly unaffected by toxic comments, many users temporarily reduce their activity or stop contributing altogether.
    }
    \label{fig:intro}
\end{figure}

Figure \ref{fig:intro}. shows the activity of 50 randomly selected users who have received exactly one toxic comment. While some users are seemingly unaffected by a toxic comment, others temporarily reduce their activity or leave the project completely. The aim of our paper is to quantify this effect on the entire population of editors.

We estimate the number of lost active days due to a toxic comment by comparing the number of active days before and after receiving a toxic comment. To account for potential baseline change, we have matched users that received a toxic comment with similarly active users who received a non-toxic comment. We have separately studied if toxic comments increase the probability of users leaving the project altogether. Finally, we have used an agent-based model to model the potential impact of an increased number of toxic comments on Wikipedia.
\section{Results} \label{results}

\begin{figure}
    \centering
    \includegraphics[width=0.8\hsize]{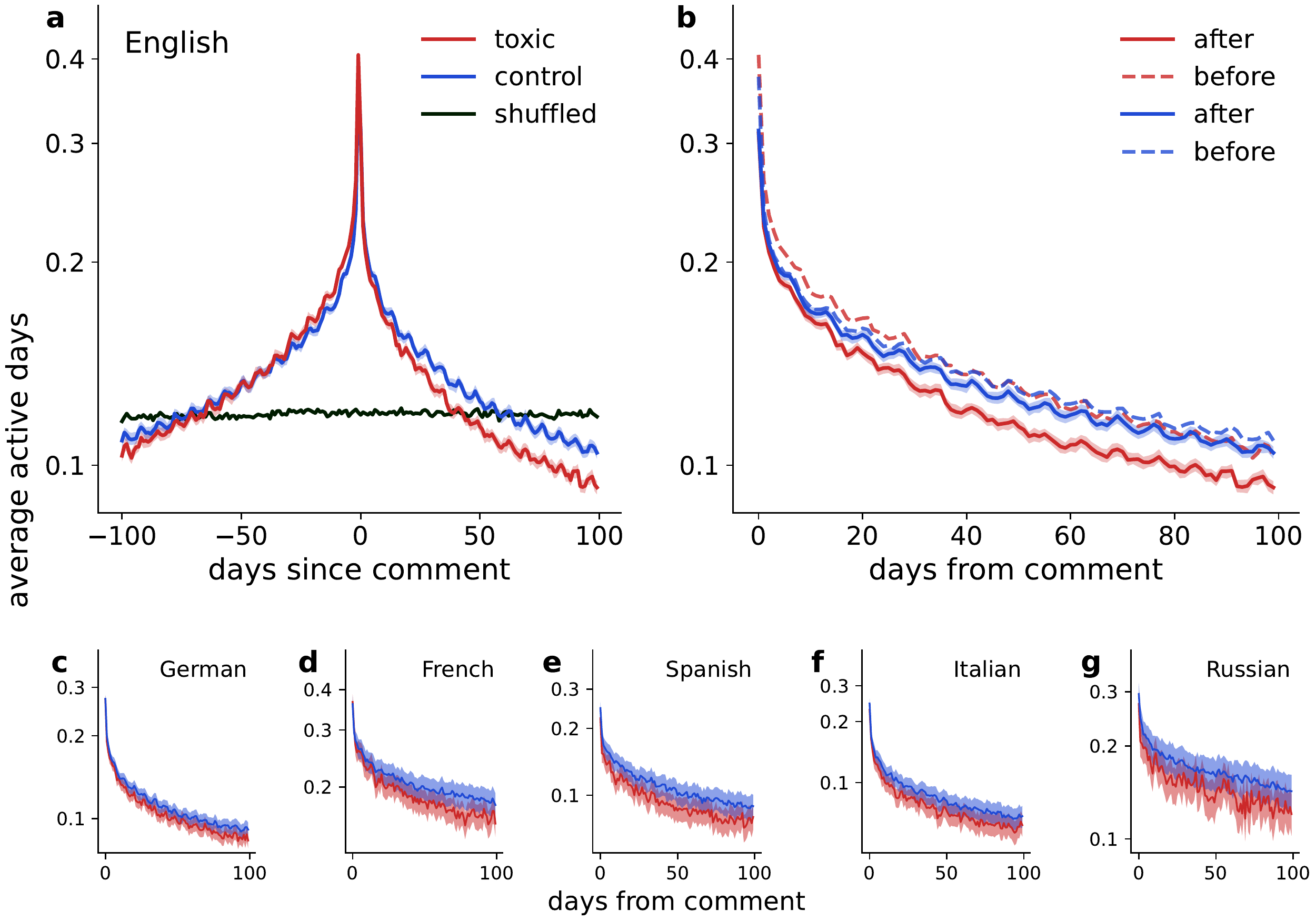}
    \caption{
        \textbf{After receiving a toxic comment, editors become less active.
        }
         On average, users are more active near the time when they receive a toxic comment (peak at zero for the red line in panel \textbf{a}). 
        Average activity across all users who have received a toxic comment is lower in all $100$ days after the event compared to the corresponding days before (dashed and solid red lines in panel \textbf{b}). This cannot be explained by a baseline drop in activity after a non-toxic comment (dashed and solid blue lines in panel \textbf{b}). Similar results hold not only for the English edition but also for the other five editions (\textbf{c-g}). 
    }
    \label{fig:main}
\end{figure}

\subsection{Loss of editor activity}
To estimate the effect of a toxic comment, we compute the proportion of users who were active on day X before or after receiving a toxic comment (Figure \ref{fig:main}). We find that, on average, editors are more active near the time when they receive a toxic comment, with a peak at $24$ hours prior to the comment. At this time point, more than $40\%$ of editors were active, as shown by the red line in Figure \ref{fig:main}a. This is a rather unsurprising observation since toxic comments are often made as a reaction to an edit made by a user and, thus, users are expected to be active around the time of a toxic comment. Note that if the timestamps around which the curve is centered are shuffled (black line in Figure \ref{fig:main}a) then this pattern disappears completely as expected. 

We also find that average activity across all users who have received a toxic comment is lower during all 100 days after the event compared to the corresponding days before (dashed and solid red lines in Figure \ref{fig:main}b), e.g. smaller number of users is active five days after receiving a toxic comment than five days before receiving it. To rule out the possibility that this is due to a general drop in activity over time or a drop in activity after any comment, we select a control group of users who have received a non-toxic comment, and whose average activity in the $100$ days before the comment is the same as the average activity of users who received a toxic comment (see \nameref{methods} for details).

We observe a similar characteristic peak around the non-toxic comment, likely due to both toxic and non-toxic comments being reactions to a contribution made by an editor. However, in contrast to a toxic comment, a non-toxic comment does not lead to a significant decrease in activity (dashed and solid blue lines in Figure \ref{fig:main}b). Similar results hold for all six language editions that we have examined (Figure \ref{fig:main}c-g). 

We then estimate the lost activity due to a toxic comment by computing the decrease in activity after a toxic comment, taking into account a potential baseline drop, i.e. by computing $\Delta =$ (After\textsubscript{toxic} $-$ Before\textsubscript{toxic}) $-$ (After\textsubscript{non toxic} $-$ Before\textsubscript{non toxic}). We find that this loss is statistically significant for all language editions studied (Table \ref{tab:lost}). We further explored the robustness of this result with respect to the toxicity threshold and potential filtering of users according to their activity. As expected, for higher toxicity thresholds, i.e. for more severely toxic comments, the effect is stronger (Figure \ref{fig:robust}). Considering only active users also leads to higher estimates, however, here we are reporting a conservative estimate, i.e. no filtering is used for results presented in Figure \ref{fig:main} and Table \ref{tab:lost}.

\begin{table}[bt]
    \caption{\textbf{Lost active days in the $100$ days following a toxic comment}. The lost active days are estimated by computing the difference between the number of active days during $100$ days after a toxic comment and the number of active days during 100 days before a toxic comment. This difference is then compared with the baseline drop after a non-toxic comment, i.e. $\Delta =$ (After\textsubscript{toxic} $-$ Before\textsubscript{toxic}) $-$ (After\textsubscript{non toxic} $-$ Before\textsubscript{non toxic}). The $P$-value is computed using Student's $t$-test} 
    \label{tab:lost}
    \centering
    \begin{tabular}{l r r r}
    \toprule
    Edition & $\Delta$ & $P$-value & $N_{users}$ \\
    \midrule
    English & $-1.207$ & $2.6 \times 10^{-66}$ & $36,332$ \\
    German & $-0.546$ & $1.5 \times 10^{-7}$ & $10,346$\\
    French & $-1.851$ & $4.8 \times 10^{-9}$ & $2,239$\\
    Spanish & $-0.563$ & $8.6 \times 10^{-3}$ & $2,446$\\
    Italian & $-0.336$ & $2.3 \times 10^{-2}$ & $3,567$ \\
    Russian & $-1.219$ & $7.8 \times 10^{-4}$ & $1,134$ \\
    \bottomrule
    \end{tabular}
\end{table}

Note that given that thousands of users have received at least one toxic comment (Table \ref{tab:summary}), even a moderate loss per user could result in many human-years of lost productivity for Wikipedia in the short run. By multiplying the estimated loss per user from Table \ref{tab:lost} by the number of users who have received at least one toxic comment, we could estimate the total loss of activity that is ranging from $5$ human-years for Russian Wikipedia to $265$ human-years for the English edition. The reason for the lasting effect of toxicity is that some new users are discouraged by a toxic comment and choose to leave the project altogether after just a few contributions. This means that a single toxic comment could deprive Wikipedia of a potentially long-term contributor.

To further investigate this effect we compare the probability of leaving Wikipedia after receiving a toxic comment with the probability of leaving Wikipedia after receiving a non-toxic comment.

\subsection{Leaving Wikipedia}
\begin{figure}
    \centering
    \includegraphics[width=0.8\hsize]{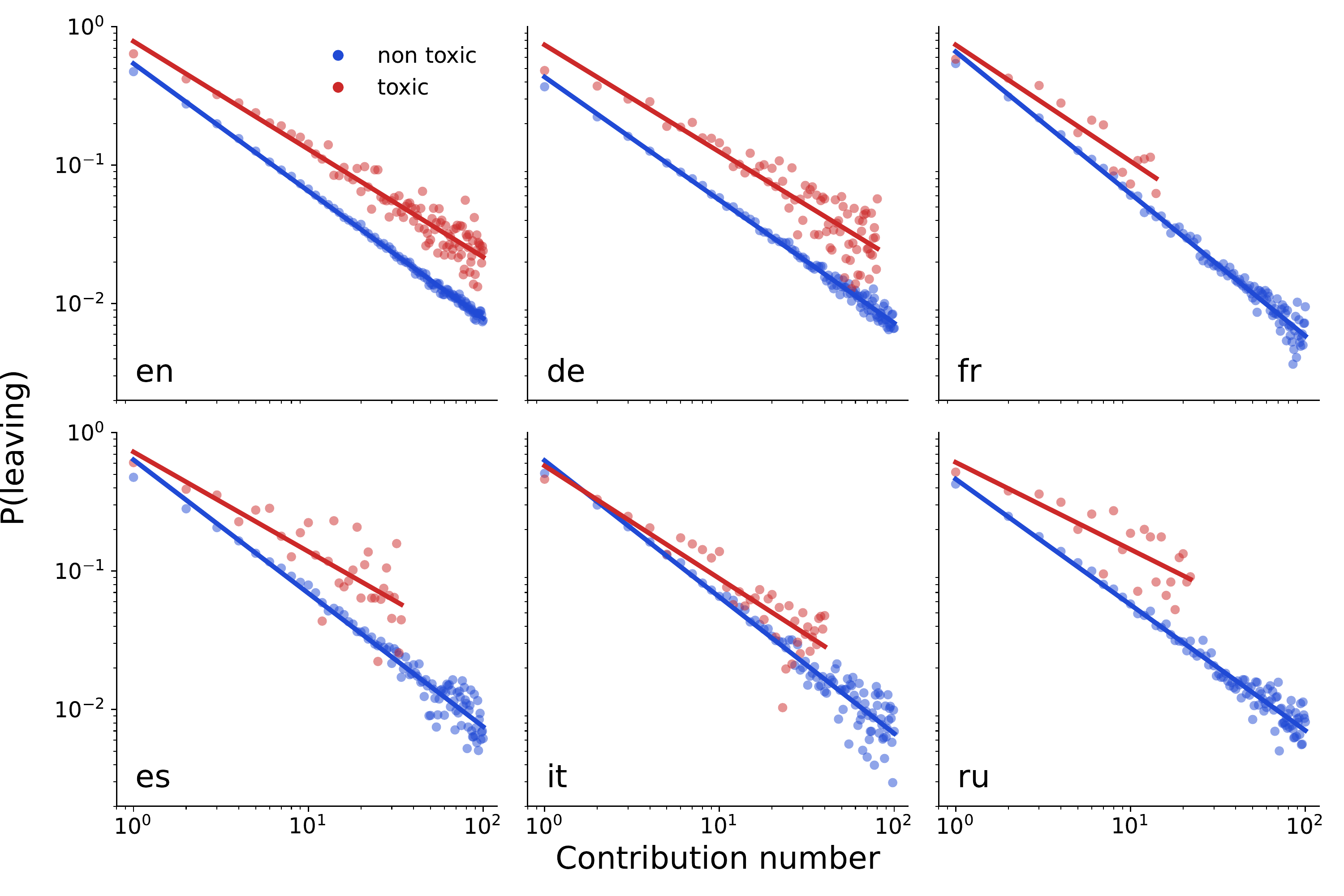}
    \caption{
        \textbf{The probability of leaving Wikipedia after receiving a toxic comment is substantially higher than might be expected otherwise.} For all six editions the probability of leaving declines with the number of contributions approximately following the power law. At the same time, this probability is substantially higher after receiving a toxic comment than might be expected otherwise. Dots are probability estimates and solid lines are the best linear fit on a log-log scale.
    }
    \label{fig:prob}
\end{figure}
We observed that the probability of leaving Wikipedia after $N$ contributions declines with $N$ following a power law. That is $P_N(\text{leaving}) \sim N^{-\alpha}$, where \textit{alpha} ranges from $0.89$ to $1.02$ for different language editions. While the probability of leaving the project after the first and only contribution is high ($P_1 = 47\%$ for English Wikipedia), the risk of leaving Wikipedia drops to $0.7\%$ for users who have made 100 contributions.

\begin{figure}[htp]
    \centering
    \includegraphics[width=0.8\hsize]{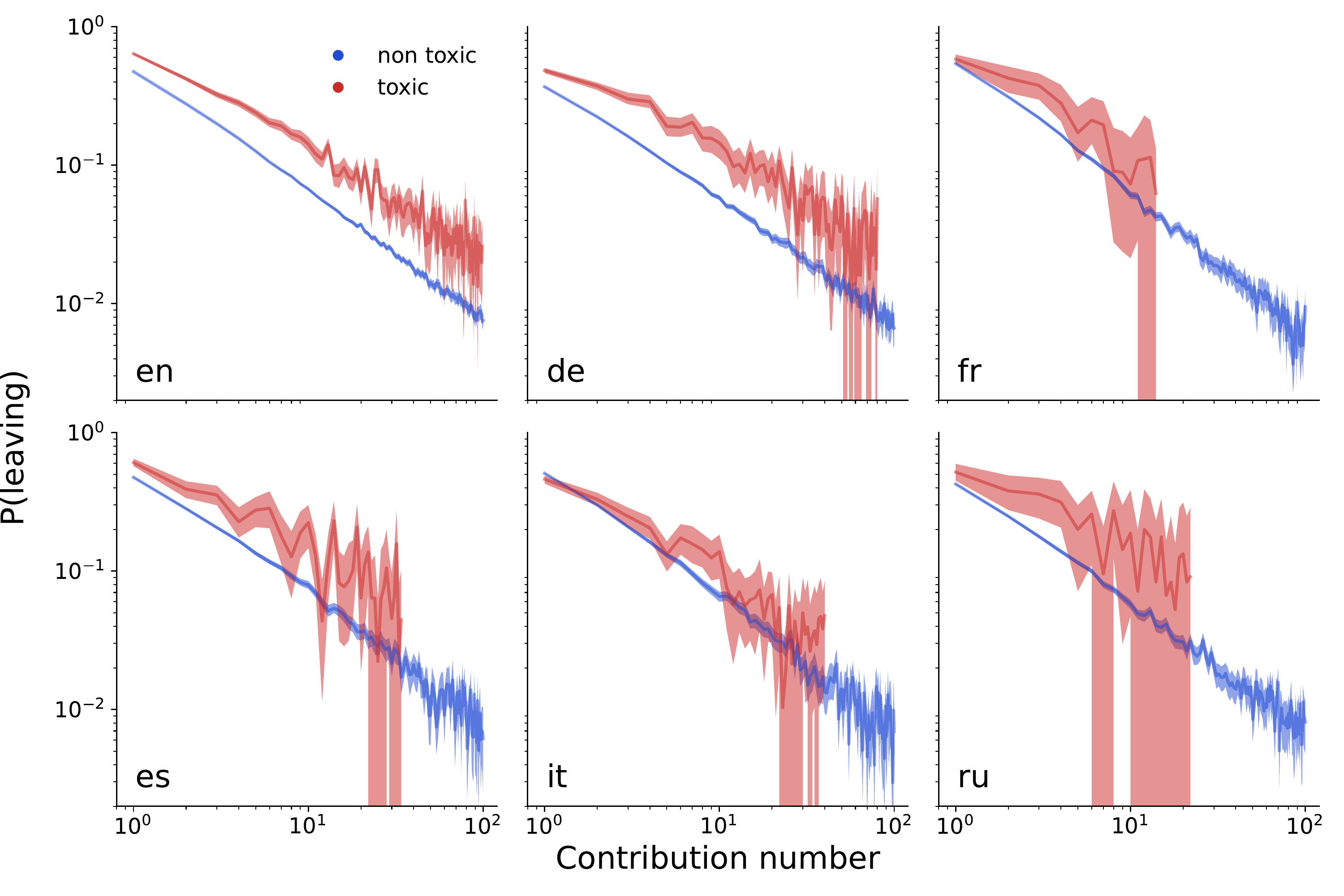}
    \caption{
        \textbf{Bootstrapped confidence intervals for the probability of leaving Wikipedia after receiving a toxic.} For the absolute majority of data points, the estimated probability of leaving is higher after a toxic comment than otherwise with no intersection between $95\%$ confidence intervals making the results significant with $P < 6.25 \times 10^{-4}$. The estimates are less reliable for higher contribution numbers because of the decreasing sample size.
        }
    \label{fig:robust_prob_sign}
\end{figure}

To study the effects of toxic comments we separately consider contributions that are followed by a toxic comment and contributions that are not followed by a toxic comment (see \nameref{methods} for details). We find that the risk of an editor leaving after a toxic comment is consistently higher for all editions and regardless of the contribution number, see Figure \ref{fig:prob}. We provide an analysis of the significance of these findings in Figure \ref{fig:robust_prob_sign}.

\subsection{Agent-based modeling}

As has been demonstrated earlier, toxic comments increase the likelihood of editors abandoning Wikipedia. If enough editors leave, this could potentially impede the progress of the project as a whole. In order to estimate the potential impact of toxic comments, we compare two environments: a non-toxic environment, where the probability of a user leaving follows empirically observed non-toxic probability distribution (blue dots in Figure \ref{fig:prob}), and a highly toxic environment, where the probability of leaving corresponds to an empirically observed toxic probability distribution (red dots in Figure \ref{fig:prob}).

To model the impact of toxic comments, we employ an agent-based approach in which different scenarios are defined by the number of users joining the project each day. Specifically, we consider the following three scenarios: 
1) $1000$ users join the project on day $1$, with no new users arriving later;
2) one new user joins the project each day indefinitely;
3) one new user joins the project each day for the first $500$ days, with no new users joining the project later.

In our model, each agent makes a contribution upon joining the project and then continues to contribute following a Poisson process, i.e in such a way that the distance between two consecutive contributions, $D$, follows an exponential distribution: $D \sim \operatorname{Exp}(\lambda)$, where $\lambda$ is estimated from empirical data. After each contribution, the editor leaves the project with the probability determined by the environment (toxic or non-toxic) and the number of contributions already made, as represented in Figure \ref{fig:prob}.

\begin{figure}
    \centering
    \includegraphics[width=1\hsize]{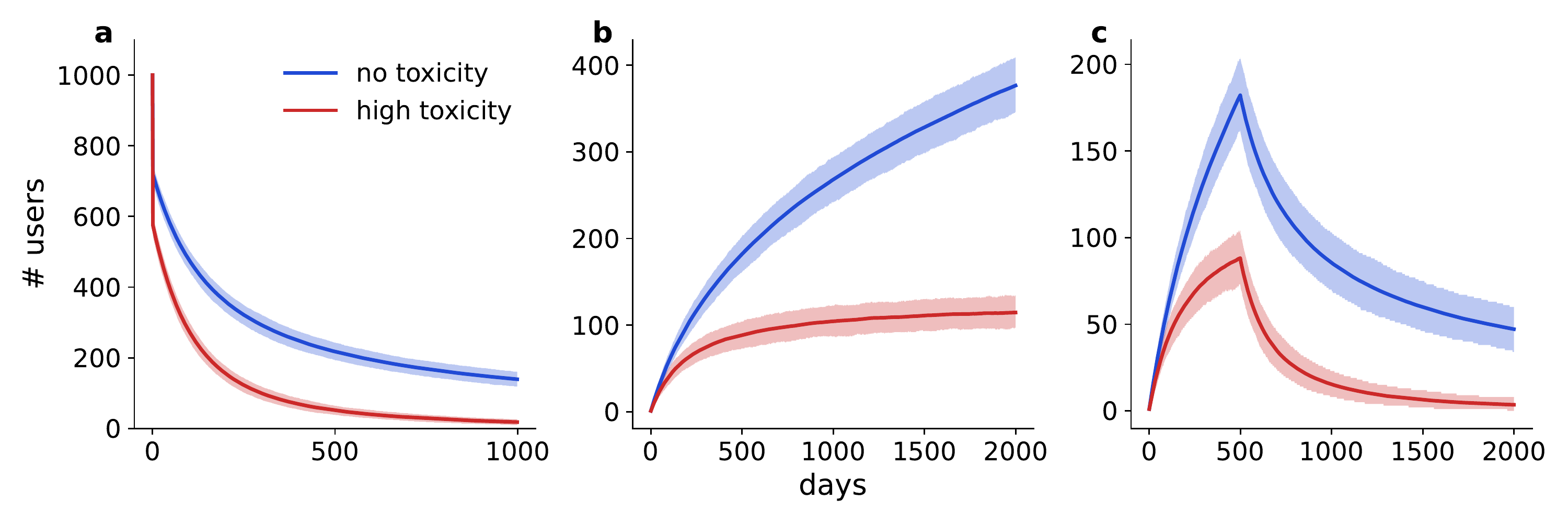}
    \caption{
        \textbf{Comparison of the number of users over time in a non-toxic vs. a toxic environment.
        }
        In a non-toxic environment, the initial population of $1000$ users (\textbf{a}) stabilizes at around $200$, whereas in a toxic environment, the population nearly dies out. If users are constantly arriving to sustain the population in a toxic environment (\textbf{b}), it would lead to growth in a non-toxic environment. When users arrive only up to a certain point in time (\textbf{c}), the peak activity of the project is substantially lower in a toxic environment compared to a non-toxic environment and activity eventually dies out.
    }
    \label{fig:model}
\end{figure}

Our model suggests that in a non-toxic environment, approximately $200$ users from the initial population of $1000$ would become long-term contributors sustaining the project. In contrast, in a toxic environment, the decline in the number of users would be faster, and almost none of them would remain in the long run, as illustrated in Figure \ref{fig:model}a

Sustaining the project in the long run in a toxic environment would require a constant influx of new users, Figure \ref{fig:model}b. The same rate of new users in a non-toxic environment would lead to the growth of the project. Figure \ref{fig:model}c illustrates a combination of the first two scenarios. The model demonstrates that toxic comments not only decrease the number of users but can also lead to qualitatively different outcomes for the project, e.g. stable vs dying out. That might be particularly important given that the actual activity on Wikipedia declines (see Figure \ref{fig:history}).

\section{Discussion} \label{discussion}
Toxic speech is a general threat to all digital platforms and communities \cite{vogels2021state}, however, it is especially dangerous to Wikipedia which relies on the contributions of volunteer editors to create articles with broad, global exposure. To ensure objectivity and comprehensiveness of articles, it is critical to establish and maintain an open and inclusive collaborative environment. However, this is not always the case on talk pages, where clashing editor opinions can lead to toxic and harmful comments threatening productivity and diversity of editors \cite{lir2021strangers,bear2016women,lenhart2016online}. It could also lead to the idea that harassment is a possible tool for silencing users.

\begin{figure}[htp]
    \centering
    \includegraphics[width=1\hsize]{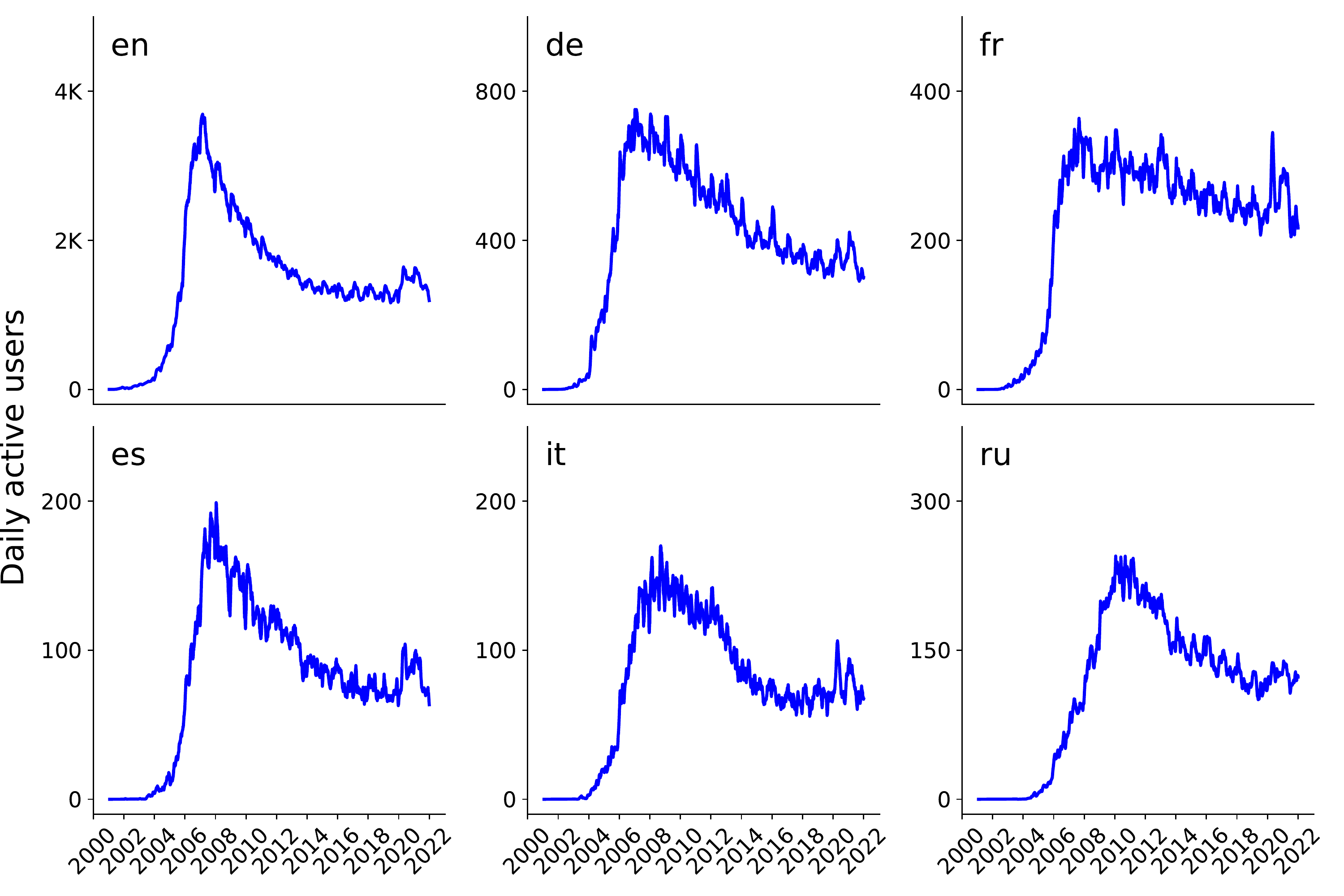}
    \caption{
        \textbf{The number of daily active users on Wikipedia over time.} After a period of exponential growth, the activity of Wikipedians slowed and then declined with an exception of a COVID-related peak in activity.
        }
    \label{fig:history}
\end{figure}

Despite that, the opinion that toxic comments are negligible and should be seen as merely over-enthusiastic participation is still present among editors \cite{corple2016beyond}. Furthermore, various anti-harassment measures have been declined multiple times by the community, as they were seen to slow the process of content creation \cite{wish1,wish2}. While multiple surveys were conducted, there existed no current analysis of toxicity impact at scale to determine whether Wikipedia should take toxic language on Wikipedia more seriously and adapt new measures to prevent further occurrences. Our large scale analysis covering all comments made on the six most active language edition of Wikipedia over the course of 20 years demonstrates that the question of toxicity should not be disregarded as it could lead to substantial loss of productivity among editors. This is especially relevant since, after a period of exponential growth \cite{almeida2007evolution}, the activity of Wikipedia editors has slowed \cite{suh2009singularity} and then declined (Figure \ref{fig:history}), with the exception of a COVID-19 related spike in activity \cite{ruprechter2021volunteer}.

Wikipedia plays a crucial role in the global information infrastructure aiming to provide millions of people with access to free, unbiased knowledge. With its reputation as a neutral and comprehensive information source, it has become a trusted first choice source of knowledge for many and its articles frequently appear in top search engine results. While it is often assumed that Wikipedia is free of toxicity, our findings demonstrate the toxicity is not only present but its impact is substantial. These findings emphasize the importance of mitigating toxic speech on collaborative platforms such as Wikipedia to ensure their continued success.
\section{Methods \& Materials} \label{methods}

\subsection{Data and Preprocessing}
\subsubsection{Comments on User's Talk Pages}
The Wikimedia Foundation provides publicly accessible dumps of all the different wikis’ content\footnote{\url{https://meta.wikimedia.org/wiki/Data_dumps} [[Accessed on 20.01.2023]]}. These dumps are updated on a regular basis, with complete revision history dumps generated once per month. For this paper, we used the English dump from 01.11.2021, the German dump from 01.08.2022, the French, Italian, and Spanish dumps from 01.08.2022, and the Russian dump from 01.07.2022. The data was obtained from a mirror hosted by the Umeå University, Sweden\footnote{\url{https://mirror.accum.se/mirror/wikimedia.org/}}. 

From the dumps, the user's talk pages were extracted. A user’s talk page is a place where other editors can communicate with the user either on more personal topics or to extend their discussion from an article’s talk page. When the comments left on the talk page are resolved or become too old, users can choose to archive them. This helps them keep better track of new incoming topics. Once archived, the old comments are not displayed on the talk page anymore but are rather linked in a subpage. Nevertheless, the entire history of the user talk page, as of any other page on Wikipedia, can be fully seen under the tab of revision history. The revision history records one entry for every edit made on the page saving each time the complete content of the page. Thus retrieving a single comment requires performing the difference between two consecutive revisions. The Wikimedia API does offer a method to compute the difference between two revisions, however, applying it on a scale that was necessary for this research was unfeasible. For that reason, we developed our own parser to extract comments as a difference between two versions of the page \cite{repo}.

We excluded from our analysis talk pages that belong to unregistered users, e.g. users who are represented only by an IP address rather than a user name, because IP addresses are dynamic and it can not be assumed that one address represents a single user throughout Wikipedia history. Additionally, we have excluded comments made by officially registered bots. Comments that were made by users on their own pages are also not considered. 

When extracting comments, we cleared wiki-specific formatting and HTML markup, i.e. removed links, attachments, or other formatting-specific sequences irrelevant to the actual content.

\subsubsection{Contributions and active days}
In order to extract information on users' contributions, i.e. edits of Wikipedia pages made by them, we used the MediaWiki API to retrieve timestamps for each edit made by a given user. The resulting data set is publicly available in the project repository \cite{repo}. The timestamps of contributions were then converted into active days. Specifically, each user $i$ was represented as a binary vector $u_i = (a_{i1}, a_{i2}, \dots, a_{iN})$, where $a_{id} = 1$ if user $i$ made at least one contribution, i.e. edited a Wikipedia page, within the 24h period corresponding to day $d$ and $a_{id} = 0$ otherwise. $N$ is the number of days between the first recorded contribution in our data set and the last. The conversion from contribution count to active days was performed because it is hard to interpret and compare the total number of contributions between users as one large contribution could be equivalent to multiple smaller ones. Additionally, the size of a contribution does not necessarily reflect the effort put into it. While being active on a given day could still mean different levels of activity for different users, it represents a certain level of engagement with the project and is substantially different from not contributing at all on a given day.

\subsection{Toxicity}
The automatic detection of offensive language in online communities has been an active area of research since at least 2010 \cite{xu2010filtering}. Over the past decade, researchers have focused on detecting closely-related and intersecting types of offensive language such as toxicity, abusive language, and hate speech \cite{fortuna2020toxic}, see \cite{zampieri2020semeval} for an overview of recent advancements in the field. In this paper, we use a model from the Perspective API \cite{perspective} to identify toxic comments. This is a state-of-the-art toxicity detection algorithm that obtained competitive results at OffensEval-2019 competition \cite{zampieri2019semeval} without any additional training on the contest data and is often used as a baseline system for toxicity detection \cite{fortuna2020toxic}. Perspective API is used across multiple platforms, including The New York Times, Der Spiegel, Le Monde, and El País. It is a BERT-based model \cite{lees2022new} trained on comments from a variety of online sources, including Wikipedia. Each comment is labeled by 3-10 crowdsourced raters. Perspective models provide scores for several different attributes, see Table \ref{tab:perspective1} for the list of attributes and their definitions, see Table \ref{tab:perspective2} for examples of toxic comments, and see Table \ref{tab:perspective3} for the AUC scores for those languages and attributes that were used in this paper.

\begin{figure}[htp]
    \centering
    \includegraphics[width=0.65\hsize]{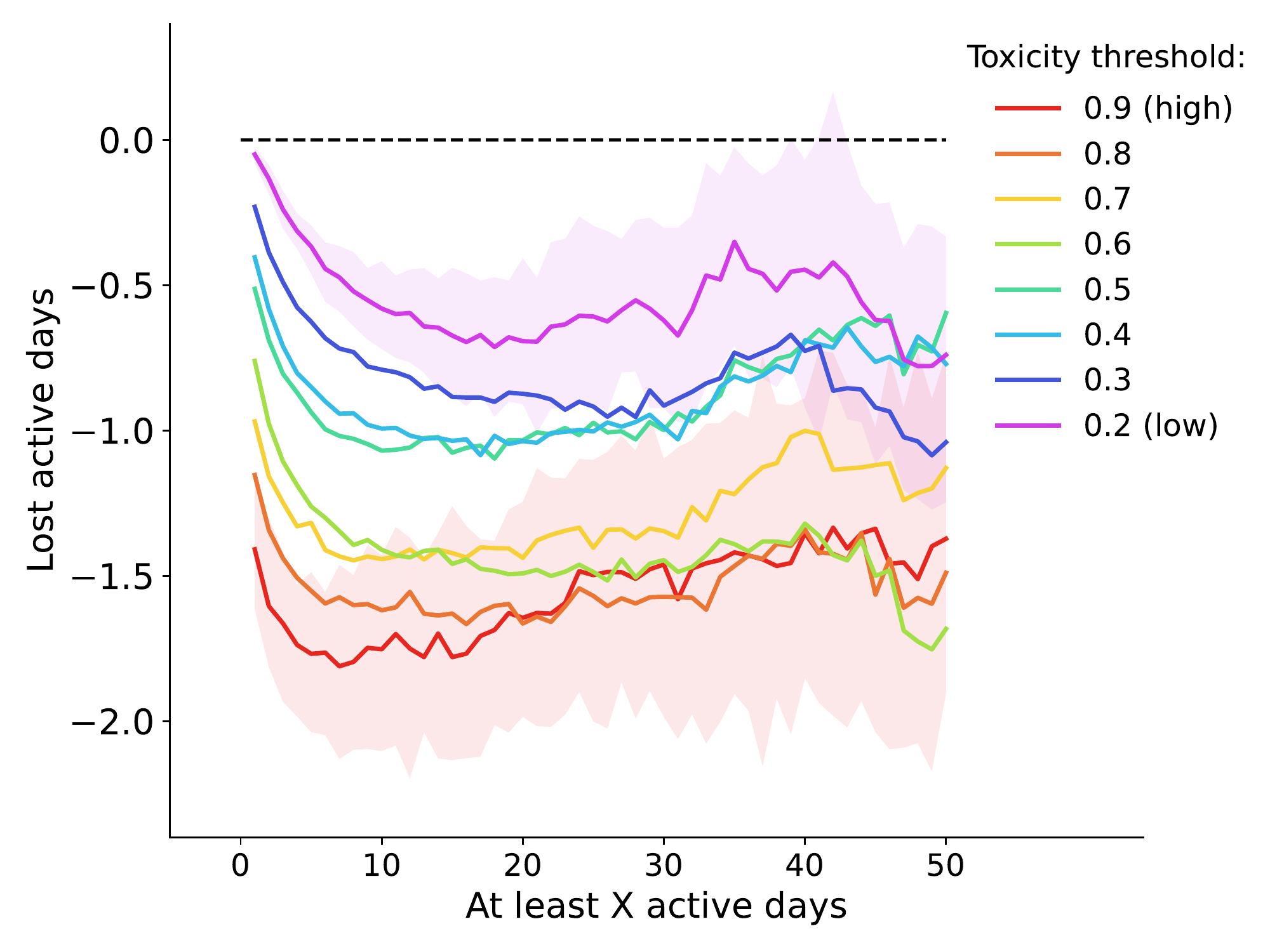}
    \caption{
        \textbf{Lost activity estimates as a function of toxicity threshold and activity level of users
        }
        We find that our results are robust with respect to toxicity threshold and filtering out less active users. For visual clarity, the 95\% confidence intervals (shaded regions) are shown only for $0.2$ and $0.9$ thresholds.
    }
    \label{fig:robust}
\end{figure}

\begin{figure}[htp]
    \centering
    \includegraphics[width=0.5\hsize]{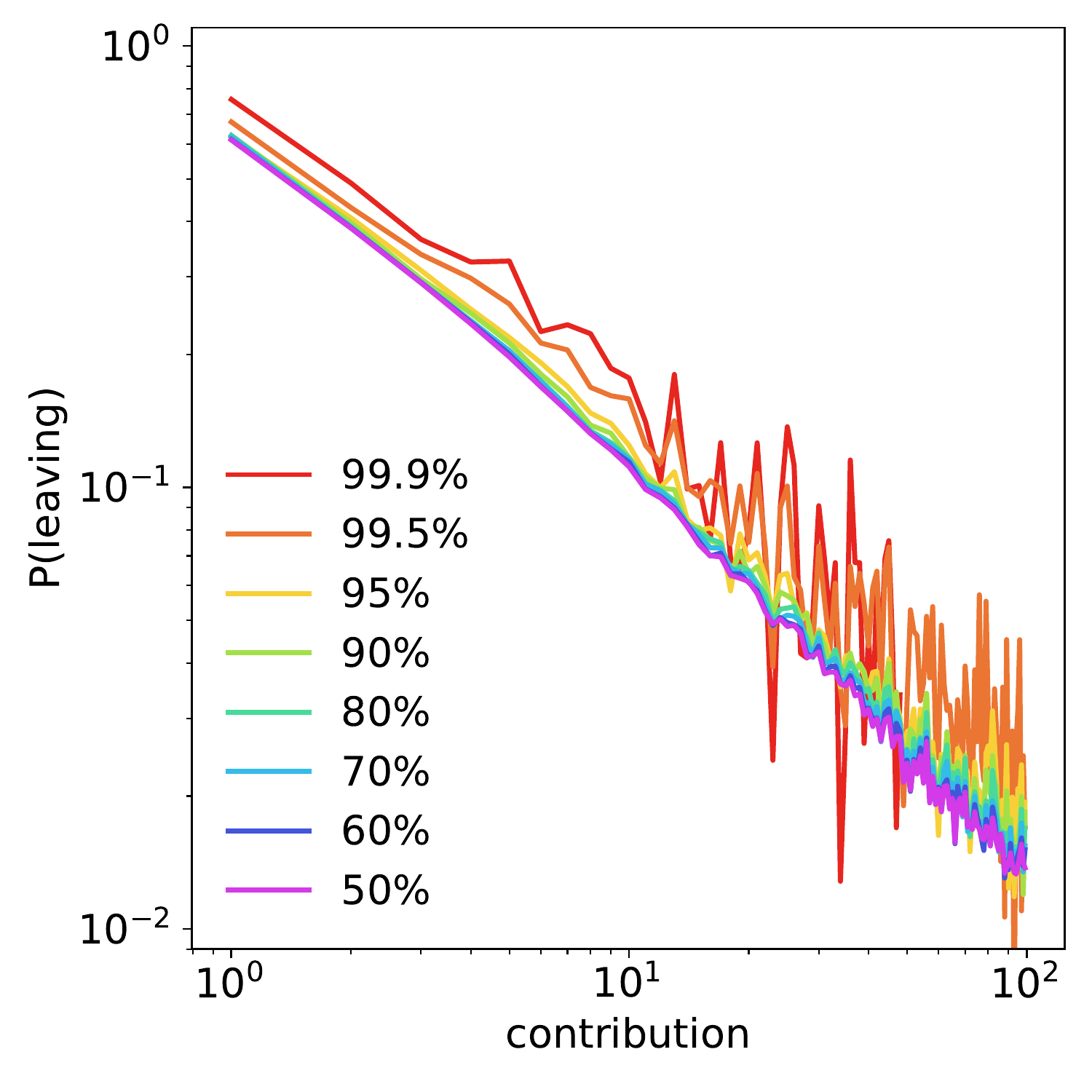}
    \caption{
        \textbf{Probability of leaving Wikipedia after receiving a toxic comment as a function of toxicity threshold}
        As expected, the probability of leaving the project is higher when higher toxicity thresholds are used. For all thresholds, this probability is substantially higher than might be expected without a toxic comment.
        }
    \label{fig:robust_prob}
\end{figure}

We define a toxic comment as a comment that has a score of at least $0.8$ on any of the six dimensions provided by Perspective API. The $0.8$ score means that on average $8$ out of $10$ raters would mark it as toxic. As this threshold can be considered arbitrary, we perform additional robustness checks using different toxicity thresholds. In particular, we compute activity loss not only for the threshold of $0.8$ (Table \ref{tab:lost}) but for thresholds from $0.2$ to $0.9$. Additionally, we applied different activity filters, e.g. we separately compute an estimate only for those users who were active at least $X$ days in the past $100$ days where $X$ varies from $0$ to $50$. This is done in order to ensure that the results are not exclusively driven by those users who had made few edits and then stopped contributing to the project. We perform this analysis for English Wikipedia as it is the largest edition. As shown in Figure \ref{fig:robust}, the estimate is typically in the range from $-0.5$ to $-2$ and significantly lower than zero for all activity thresholds and all toxicity thresholds higher than $0.3$. Similarly, we have checked how the toxicity threshold affects the probability of leaving the project. As might be expected, results remain qualitatively the same for different toxicity thresholds but higher thresholds lead to more extreme results, e.g. the probability of leaving after a toxic comment with $0.9$ score is even higher than after a toxic comment with toxicity score of $0.8$ (Figure \ref{fig:robust_prob}).

Perspective API accepts texts up to $20,480$ bytes. As the majority of comments are well below this limit, we have excluded those that are larger. 

\subsection{Activity loss}
For each user, we select a random toxic comment that they has received. We then center the user activity around the timestamp, $t_i^{tox}$, of that toxic comment, and convert the result to active days, by computing $$\operatorname{sign} (|\{t \in T_i : t \in [t_i^{tox} + d * 24 * 60 * 60, t_i^{tox} + (d + 1) * 24 * 60)\}|)$$, where $T_i$ is the set of timestamps of all contributions made by user $i$, and $d$ is a day ranging from $-100$ to $100$. Finally, the results were averaged over all users. We repeat the procedure of selecting a random toxic comment 100 times and report average results. However, as most users received only one toxic comment, there is little variation across simulations and the average over 100 simulations is almost identical to the results of a single simulation.

For comparison, the same procedure is repeated for randomly selected non-toxic comments, however, it includes an additional step as our control group should include users who are as active as those who have received a toxic comment. On average, users who have received a toxic comment are more active. This is probably due to the fact that each contribution could lead to a toxic response with a certain probability. Thus, the more contributions a user has made the higher the chance of receiving a toxic comment. To account for that, for each user from a non-toxic group we generate with a probability $p$ a virtual toxic comment after each contribution. We then select for a control group not all users but only those who have received at least one virtual toxic comment. The $p$ is selected in such a way that the average activity in $100$ days before a toxic or non-toxic comment matches. 

To test for the significance of the results we compute 95\% bootstrapped confidence intervals for each estimate.

\subsection{Probability of leaving}
For each toxic comment, we find the closest in time contribution that precedes that comment. We define such contributions as "contributions followed by a toxic comment" and compare the probability of leaving after such contributions with the probability of leaving after other contributions. The probability of leaving after $N$ contributions is estimated as a fraction of users who have made exactly $N$ contributions among users who have made at least $N$ contributions. As the probability of leaving strongly depends on $N$, we make a comparison separately for each contribution number $N \in [1, 100]$. For $N > 100$ the number of users is too small to provide reliable estimates for comparison.
\subsection{Acknowledgment}
We acknowledge the Master's thesis by Sebastian Brückner \cite{bruckner2021modeling}, which identified a potential pattern in data and provided an inspiration for the design of the study presented in this paper.

The initial data collection and experiments were carried out as part of Camelia Oprea's Master's thesis \cite{oprea2022determining}.

We thank Liubov Tupikina and David Garcia for their valuable discussions regarding the results presented in this article.

This preprint was created using the LaPreprint template (\url{https://github.com/roaldarbol/lapreprint}) by Mikkel Roald-Arb\o l \textsuperscript{\orcidlink{0000-0002-9998-0058}}.

\bibliography{main}


\clearpage


\section{Supplementary Material} 
\supplementarysection

\begin{table}[htp]
    \small
    \caption{Summary statistics for the six language editions used in the study} 
    \label{tab:summary}
    \begin{tabular}{l r r r}
    \toprule
    Edition & Comments & \multicolumn{2}{c}{Users with at least}  \\
     &  & one comment & one toxic comment\\
    \midrule
    English & $37,304,436$ & $6,216,906$ & $80,307$ \\
    German  &  $8,015,691$ &   $659,167$ & $19,543$ \\
    French  &  $3,838,161$ &   $455,598$ &  $4,282$ \\
    Spanish &  $2,964,606$ &   $525,443$ &  $5,972$ \\
    Italian &  $2,816,537$ &   $395,257$ &  $7,666$ \\
    Russian &  $2,408,401$ &   $326,738$ &  $1,687$ \\
    \bottomrule
    \end{tabular}
\end{table}

\begin{table}[htp]
    \small
    \caption{Attributes used by Perspective API} 
    \label{tab:perspective1}
    \begin{tabular}{l p{10cm}}
    \toprule
    Attribute name & Description \\
    \midrule
    Toxicity        & A rude, disrespectful, or unreasonable comment that is likely to make people leave a discussion.\\
    Severe toxicity & A very hateful, aggressive, disrespectful comment or otherwise very likely to make a user leave a discussion or give up on sharing their perspective. This attribute is much less sensitive to more mild forms of toxicity, such as comments that include positive uses of curse words. \\
    Identity attack & Negative or hateful comments targeting someone because of their identity. \\
    Insult          & Insulting, inflammatory, or negative comment towards a person or a group of people. \\
    Profanity       & Swear words, curse words, or other obscene or profane language. \\
    Threat          & Describes an intention to inflict pain, injury, or violence against an individual or group.\\
    \bottomrule
    \end{tabular}
\end{table}

\begin{table}[htp]
    \small
    \caption{Performance of Perspective API models (AUC scores)} 
    \label{tab:perspective3}
    \begin{tabular}{l r r r r r r}
    \toprule
      & English & German & French & Spanish & Italian & Russian\\
    \midrule
    toxicity        & $0.97$ & $0.94$ & $0.94$ & $0.94$ & $0.95$ & $0.91$ \\
    severe toxicity & $0.98$ & $0.98$ & $0.96$ & $0.96$ & $0.97$ & $0.95$ \\
    identity attack & $0.97$ & $0.94$ & $0.96$ & $0.95$ & $0.96$ & $0.94$ \\
    insult          & $0.97$ & $0.93$ & $0.94$ & $0.93$ & $0.96$ & $0.92$ \\
    profanity       & $0.99$ & $0.97$ & $0.98$ & $0.98$ & $0.98$ & $0.96$ \\
    threat          & $0.99$ & $0.95$ & $0.99$ & $0.98$ & $0.98$ & $0.96$ \\
    \bottomrule
    \end{tabular}
\end{table}

\begin{table}[htp]
    \small
    \caption{Examples of comments with different toxicity scores} 
    \label{tab:perspective2}
    \begin{tabular}{l p{10cm}}
    
    \toprule
    Toxicity score & Text \\
    \midrule
    $0.07$ &
    I wouldn't trust me on this. I suggest asking [user name], who usually goes around correcting my mistakes. Also, you appear to be vastly more qualified than me. However I would have thought that geometric isomerism would be independent of the isotopic mass - the properties would be very similar (identical?). \\
    $0.14$ & Your home page doesn't work. There is nothing in your talk section either! \\
    $0.21$ & I wanted to let you know that I just tagged [[article name]] for deletion, because it seems to be vandalism or a hoax. If you feel that the article shouldn't be deleted and want more time to work on it, you can contest this deletion, but please don't remove the speedy deletion tag from the top. \\
    $0.35$ & What the heck did you remove the [category name] from articles? All that happened was that some user renamed it to remove a [word] from the name, but didn't set up a redirect. Now I'm going to have to go through all the articles that were in the category and reinsert it again. Could you not have done a quick search first just to make sure that the name of the category hadn't been slightly renamed? Sorry if I sound a bit grouchy but I have spent most of this week doing nothing but  fixing other people's screw-ups and I'm getting fed up having to do it. \\
    $0.49$ & I don't know who you think you are but I do NOT appreciate hours of work being undone. I don't know what gives you the right to undo other peoples work. You claim I am editing warring yet it feels like quite the opposite for me. I contribute to wikipedia and you trash it. Yeah, I'm not edit warring, buddy. \\
    $0.55$ & Hope you're proud of turning Wikipedia into a vehicle for attacking the reputations of real people whose biggest error was attempting to contribute and whose second biggest error was thinking that Wikipedia policies are sane. I'm not an edit warrior, like you, so I'm going to allow others to remove the libel. Feel free to edit war with them, too, and then pat yourself on the back for "defending Wikipedia against vandalism". The war on vandalism, like the war on drugs or terror, is a pointless and transparent fascist power grab. \\
    $0.62$ & Given the vile judgemental filth you have been posting in the [article name] name debate, you are not only a hypocrite; but anything I have said to you is remarkably restrained. It is clear that your only motives are to troll and to push a truly perverted point-of-view. \\
    $0.72$ & People have bent over backwards to accomodate you, mostly out of a pure concern for fairness and justice. You have eroded all the good will left in my body. Any more crap out of you, and I will "take steps". Better listen up good, because I'm one the oldest "old hands" around here and I know how to work the system. Don't get me riled up against you, or I will force you out!!  \\
    $0.83$ & I would like to thank you for the personal insult to my intelligence; as opposed to addressing the arguments in my comment. I know actually engaging in dialogue would have lowered you to my level. But the rest of the public doesn't care too much for experts who believe that the rest of the public is below them, though I doubt you are an expert on anything but [vulgar speech] and being a self satisfied [expletive] who has nothing better to do than insult people on wikipedia. \\
    $0.99$ & Ahh shut [expletive 1] up you [expletive 2] [ethnic slur]. [expletive 1] you [ethnic slur] i will find u in real life and slit your throat.\\
    \bottomrule
    \end{tabular}
\end{table}

\clearpage

\end{document}